\documentclass[fleqn,twoside]{article}
\usepackage{espcrc2}
\usepackage{graphicx}
\usepackage{epsfig}
\usepackage[figuresright]{rotating}

\newcommand{\Ztautau}{Z\rightarrow\tau\tau}

\newcommand{\Zee}{Z\rightarrow ee}

\newcommand{\DO}{D\O{} }

\newcommand{\ppbar}{p\overline{p}}

\newcommand{\Eslash}{\mbox{$\rm E \kern-0.6em\slash$}}
\newcommand{\etmiss}{\mbox{$\rm \Eslash_{T}\!$}}

\newcommand{\AmS}{{\protect\the\textfont2
  A\kern-.1667em\lower.5ex\hbox{M}\kern-.125emS}}

\title{Identification of $\tau$ leptons at the \DO experiment}

\author{R. Madar\address[MCSD]{CEA/DSM/IRFU/SPP Saclay, FRANCE} for the \DO Collaboration
}
       
\begin{document}

\begin{abstract}
The article describes the identification of hadronically decaying $\tau$ leptons in $\ppbar$ collisions at $1.96$~TeV collected by the \DO detector at the Fermilab Tevatron. After a brief description of the motivations and the challenges of considering $\tau$ leptons in high energy hadronic collisions, details of the $\tau$ reconstruction and identification will be discussed. The challenges associated for $\tau$ energy measurements in an hadronic environment will be presented including approaches to deal with such measurements.
\vspace{1pc}
\end{abstract}

\maketitle

\section{MOTIVATIONS AND CHALLENGES}

Leptons are of particulary interest in high energy hadronic collisions because they provide clean event signatures in a complex hadronic environment. Electrons and muons are often considered because of such detector signatures. In addition, $\tau$ leptons (called simply $\tau$ here after) brings a potential acceptance gain of $50-300$ $\%$ according to the lepton multiplicity of the analyzed final state. Moreover, studies with $\tau$ leptons provide an interesting area to test the consistency of the Standard Model (SM) through lepton universality at high energies by measuring the branching ratio of electroweak (EW) bosons decaying into $\tau$ leptons~\cite{ZtautauXsec}. One promising candidate for extension to the SM is the Minimal Supersymmetric Standard Model (MSSM) which predicts new particles that can decay into $\tau$ leptons. The existence of such scenario can be probed within $\tau$ final states~\cite{SquarkSearch,ChargedHiggs}. Finally, many decay chains initiated by the Higgs boson involve $\tau$ leptons and considering them can then increase the experimental sensitivity to understand the origin of electroweak symmetry breaking (EWSB)~\cite{SMHiggs}.

Considering $\tau$ leptons thus appears to encompass many virtues. However, understanding such an object in a hadronic environment is very challenging. The neutrino(s) from $\tau$ final states escape the detector without interacting and forces a fraction of the $\tau$ energy to be invisible. Hence, the resulting visible energy is relatively soft and includes contamination from soft QCD interactions, which becomes a primary component of the hadronic final state. Further, the various decay mode of $\tau$ must be considered. On the one hand, the identification of $\tau$ from its leptonic decay ($\mathcal{BR}\sim 35\%$) suffers from electrons and muons originating from direct EW bosons decay as well as a poor statistic due to $\mathcal{BR}(\tau\tau\to e\mu)=6\%$. On the other hand, $\tau$ identification from its hadronic decays ($\mathcal{BR} \sim 65\%$) have different detector signature according to the hadronic resonance involved. Moreover, direct QCD interactions from hadrons collisions produce a lot of hadronic final states which can mimic the hadronically decaying $\tau$. To be optimal, each of the different hadronic $\tau$ final states requires dedicated analysis and a combination is performed to extract the full information.
 
The complexity of hadronic $\tau$ final states require sophisticated algorithms based on multivariate technics applied after event reconstruction. The next section describes the $\tau$ reconstruction at the \DO experiment. The subsequent section discusses multivariate analysis (MVA) discriminant, which allows separation of $\tau$ from jet fakes.
\section{TAU RECONSTRUCTION}

 In this section, we describe how the $\tau$ candidate is built from elementary reconstructed objects within the calorimeter and tracking systems~\cite{DOdetector}.

\subsection{Definition of reconstructed objects}
 
Three objects are potentially used to build reconstructed $\tau$ object:
\begin{enumerate}
 \item calorimeter cluster found by a Simple Cone Algorithm in a $\Delta R\leq 0.5$ cone.
 \item electromagnetic subcluster found using a Nearest Neighbour Algorithm with a seed in the third layer of the electromagnetic calorimeter (i.e. the layer containing the finest segmentation). The clustered energy is required be greater than $800$~MeV.
 \item All the tracks found by the \DO tracking algorithm in a $\Delta R\leq 0.3$ cone around the calorimeter cluster are considered. An invariant mass requirement is applied to keep only tracks compatible with the $\tau$ decay.
\end{enumerate}

In the next section we discuss the use of these objects to optimize the $\tau$ reconstruction according to its hadronic decay.
 
\subsection{Definition of type candidate}
The \DO experiment bases its $\tau$ lepton reconstruction on specific signatures of the hadronic system present in the $\tau$ final state. Three types of candidate are defined and depends on the reconstructed objects signature. There are categorized as:

\begin{itemize}
 \item type $1$: one calorimeter cluster and one track (optimized for $\tau^{\pm}\to\pi^{\pm}\nu_{\tau}$ decay);
 \item type $2$: one calorimeter cluster, one track and one calorimeter EM subcluster 
       (optimized for $\tau^{\pm}\to\rho(\to\pi^{0}\pi^{\pm})\nu_{\tau}$ decay);
 \item type $3$: one calorimeter cluster, at least two tracks 
       (optimized for $3$-prong $\tau^{\pm}\to\pi^{\pm}\pi^{\mp}\pi^{\pm}\nu_{\tau}$ decay).
\end{itemize}

\subsection{Reconstruction efficiency}

\begin{figure}[htb]
 \vspace{9pt}
 \epsfig{file=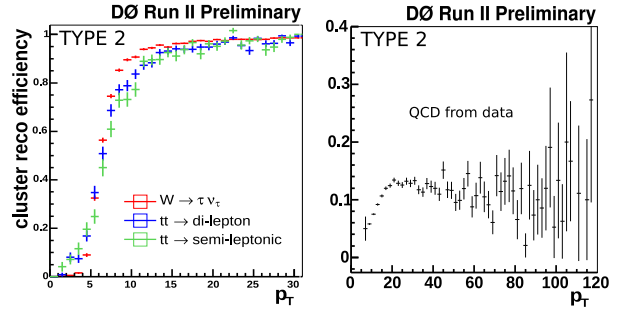,width=0.5\textwidth}
 \caption{Reconstruction efficiency for type $2$ candidates versus its transverse momentum. Left: candidates are genuine $\tau$ coming from different  simulated processes. Right: candidates are jets from an enriched QCD data sample.}
 \label{fig:recoeff}
\end{figure}

Figure \ref{fig:recoeff} shows the reconstruction algorithm for type $2$ $\tau$. One can see that almost all real $\tau$ above $20$~GeV are selected by the \DO reconstruction but a large fraction of jets (or fake $\tau$) remains. Indeed, jets from direct strong interaction can easily mimic hadronically decaying $\tau$ lepton and therefore, sophisticated methods is needed to separate $\tau$ from jets. This is the subject of the next section.

\section{DISCRIMINATION FROM JETS}

\subsection{Problematics and strategy}

\begin{figure}[htb]
 \vspace{9pt}
 \epsfig{file=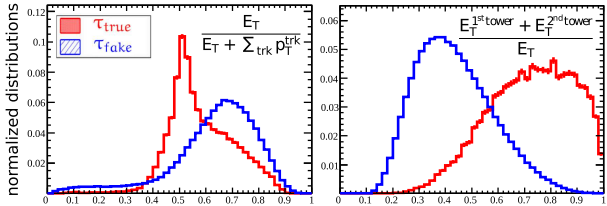,width=0.5\textwidth}
 \caption{Example of two discriminating observables used in the NN to separate jets from $\tau$. Left: the observable describes which energy fraction is taken by the neutral or the charged component of hadronic final state. Right: the observable describes if the deposited energy is spread over several calo towers: $\tau$ tend to have spread deposit to jets.}
 \label{fig:varNN}
\end{figure}

Hadronic final states of $\tau$ leptons can be faked by jets arising from direct QCD interactions. However, jets signatures from $\tau$ are narrower and have lower tracks multiplicity. These properties can be used to reject QCD jets and thereby increase the $\tau$ identification efficiency. In fact, other specificities of hadronic $\tau$ decays can also be exploited and the full information can be combined in a Neural Network (NN)~\cite{NNtechnics}. In particular, a total of $12$ observables (depending on the $\tau$ type) are considered based on a)the isolation in the tracking system as well as in the calorimeter b)shower shape and composition c)correlations between the tracking system and the calorimeter. Figure \ref{fig:varNN} provides an example of two such NN input variables.

\begin{figure}[htb]
 \vspace{9pt}
 \centering
 \epsfig{file=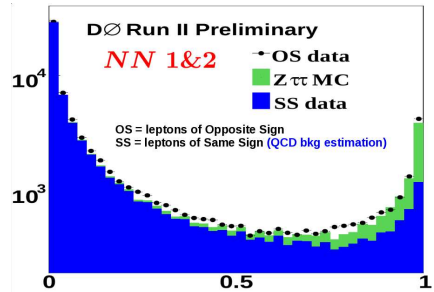,width=0.35\textwidth}
 \caption{The NN output for type $1$ and $2$ candidates in $\mu + \tau$ final states. The green histogram is the $\Ztautau$ simulation (peaking at $1$ as expected), the blue histogram is an estimation of QCD background and the black dots are the data.}
 \label{fig:NNoutput}
\end{figure}

The NN output peaks at $0$ for the QCD jets and at $1$ for the $\tau$ jets as shown in figure \ref{fig:NNoutput}. The $\Ztautau$ electroweak process is clearly sizeable and data (black dots) are well understood by the SM prediction (filled histograms).

\subsection{Identification efficiencies}

Although $\tau$ reconstruction algorithm is efficient, the signal over background ratio $(S/B)$ is poor. In fact, the order of magnitude of $S/B$ is $2$ at the reconstruction level and reaches almost $70$ after the NN selection. Table \ref{table:1} summarizes efficiencies for each $\tau$ types after reconstruction and subsequent NN identification.

\begin{table*}[htb]
\caption{Fraction (\%) of jets and $\tau$ passing the reconstruction and a NN selection for each $\tau$ types~\cite{TauEff}.}
\label{table:1}
\newcommand{\m}{\hphantom{$-$}}
\newcommand{\cc}[1]{\multicolumn{1}{c}{#1}}
\renewcommand{\tabcolsep}{2pc} 
\renewcommand{\arraystretch}{1.2} 
\begin{tabular}{@{}lllll}
\hline
$\tau$ type & \cc{$1$} & \cc{$2$} & \cc{$3$} & \cc{all}\\
\hline
jets after reco. only       & \m2  & \m12 & \m38 & \m52 \\
$\tau$ after reco. only     & \m11 & \m60 & \m24 & \m95 \\
\hline
jets  having NN$\geq0.9$      & \m0.06  & \m0.24  & \m0.80 & \m1.1 \\
$\tau$ having NN$\geq0.9$     & \m7     & \m44    & \m16   & \m67  \\
\hline
\end{tabular}\\[2pt]
\end{table*}

\subsection{Further optimizations}

The optimization strategy is based on how the NN treats the information provided by the physics. By denoting $\vec{X}=(x_1,x_2,..,x_n)$ as a point in the discriminating variables space, the NN output $\eta^{nn}(\vec{X})$ converges to the true likelihood function (which is the best classifying function):
\begin{equation}
 \eta^{true}(\vec{X}) \equiv \frac{\mathcal{S}(\vec{X})}{\mathcal{S}(\vec{X}) + \mathcal{B}(\vec{X})}
\end{equation}

\noindent Here, $\mathcal{S}(\vec{X})$ and $\mathcal{B}(\vec{X})$ are the probability density function of the genuine $\tau$ (signal) and QCD jets (background) respectively in the discriminating observables space.

One can find two strategies to optimize the separation between jets and $\tau$. A first approach is to improve $\eta^{true}(\vec{X})$ by adding more observables based on physical properties of $\tau$ leptons. A separate method would be to minimize $|\eta^{true} - \eta^{nn}|$ by performing the NN training in specific phase space regions or by increasing the statistics of the training samples to describe the differences in more details.

\begin{figure}[htb]
 \vspace{9pt}
 \centering
 \epsfig{file=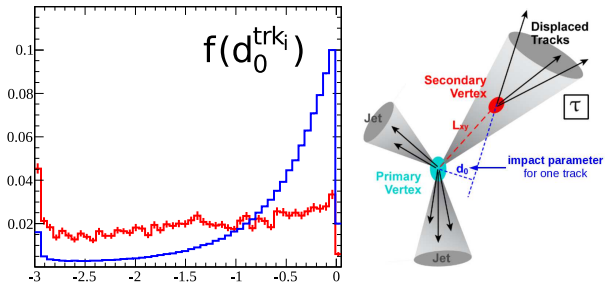,width=0.5\textwidth}
 \caption{Left: signal and background distribution of an observable based on the track impact parameter (in blue for the QCD jets and in red for the real $\tau$). Right: a scheme of experimental signature of long lived particle.}
 \label{fig:NNbtagvar}
\end{figure}

In practice, the first approach was tested by trying to include information from the \DO preshower system~\cite{DOdetector} which has a better segmentation than the calorimeter and allows an improved separation of the $\pi^0$ from the $\pi^{\pm}$ (for type $2$ decay). After several studies, it has been shown that this approach does not significantly help to separate $\tau$ from QCD jets. New physical information can be used such as the long lifetime of the $\tau$ lepton which can lead to displaced tracks from the primary vertex similar to the case of b-jets. Figure \ref{fig:NNbtagvar} shows a new discriminating variable based on the impact parameter of each track for type $3$ $\tau$ candidates. This new observable increases the $\tau$ identification efficiency of about $10\%$ for the same jet rejection.

A second approach consisting in helping the NN convergence was applied by using larger training samples and tuning specific NN parameters such as the numbers of nodes, epochs and minimization algorithm~\cite{NNtechnics}. Moreover, the difference between $\tau$ and QCD jets can evolve with the candidate transverse energy. In order to help the NN to exploit this behaviour, dedicated trainings were done for events with high ($\geq 45$~GeV) and low ($\leq 45$~GeV) $p_T$  $\tau$ candidate.

\begin{figure}[htb]
 \vspace{9pt}
 \centering
 \epsfig{file=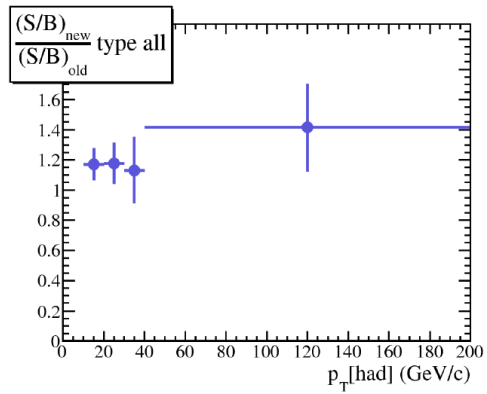,width=0.40\textwidth}
 \caption{Impact of optimizations on hadronically decaying $\tau$ leptons identification versus the $p_T$ candidate. $S/B\equiv N(\tau)/N(jets)$ and the mention ``old''  (resp. ``new'') refers to an event selection based on NN without (resp. with) optimizations. The relative gain is around $20\%$ and is better at high $p_T$ as expected thanks to the specific training.}
 \label{fig:NNptdep}
\end{figure}

Overall, the $\tau$/jet discrimination is improved by $15\%$ at low $p_T$ and $40\%$ at high $p_T$ as shown in figure \ref{fig:NNptdep}.

\section{TAU ENERGY MEASUREMENT}

In order to perform physical measurements in $\tau$ lepton final states, the identification of $\tau$ leptons is necessary but not sufficient. One needs to have a good understanding of $\tau$ kinematic properties. This section deals with the $\tau$ energy measurement at the \DO experiment.

\subsection{Problematics and strategy}
The object energy calibration at collider experiments is usually performed with the help of well known physical processes. At high energies, the production and decay of $Z$ bosons allow to perform such calibrations. Two important challenges arise for the hadronically decaying $\tau$. Firstly, the $\Ztautau$ visible mass\footnote{The visible mass observable for $\Ztautau$ events is defined by $M^2_{vis} = (p_{\tau_1} + p_{\tau_2} + \etmiss)^2$, where $\etmiss$ is the transverse missing energy.} peak is broad and shifted by the neutrinos present in the $\tau$ decay which makes this observable less sensitive than for the $\Zee$ decay. Secondly, the various $\tau$ decay modes implie lower statistics compared with the $\Zee$ decay ($\sim 10^{3}$ events versus $\sim 10^{5}$ events). These situations require to find another strategy to calibrate the visible energy of hadronic $\tau$ lepton.

If one considers a type $2$ candidate having visible energy from $(\pi^{0},\pi^{\pm})$, the true energy can be obtained from the measured energy using the hadronic calorimeter response ($R_{\pi}$) and the electromagnetic calorimeter response $(R_{e})$:
\begin{equation}
 E^{true} = R_{\pi} E^{meas}_{\pi^{\pm}} + R_{e} E^{meas}_{\pi^{0}}
\end{equation}

However, the \DO calorimeter is not compensated ($R_{e} \neq R_{\pi}$) and its segmentation does not allow to separate the $\pi^{\pm}$ shower from the $\pi^{0}$ one. Thus, the only measurable energy is $E^{meas}_{\pi^{\pm}} + E^{meas}_{\pi^{0}}$. In order to deal with these experimental constraints, one must use the track energy as reference and  propagate it to the calorimeter in two ways: the absolute correction and the relative correction. Both methods are described below.

\subsection{Absolute correction}
The idea of the absolute correction is to use the tracker to measure the $\pi^{\pm}$ energy. In order to avoid double counting, the average of the energy deposited by charged pion in the calorimeter is subtracted:
\begin{equation}
 E^{corr} = E^{trk} + E^{cal} - \langle R_{\pi}(E^{trk},\eta)\rangle \cdotp E^{trk}
\end{equation}

\begin{figure}[htb]
 \vspace{9pt}
 \centering
 \epsfig{file=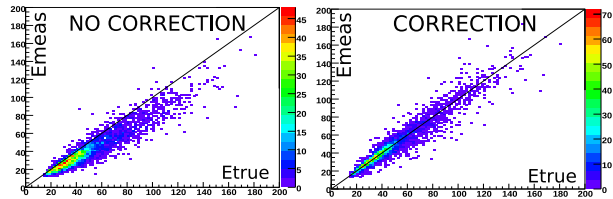,width=0.5\textwidth}
 \caption{The measured visible energy versus the true energy with (right) and without (left) absolute correction. This study done on simulated events show that the absolute correction give a better energy measurement.}
 \label{fig:EnergyCorrAbs}
\end{figure}

Figure~\ref{fig:EnergyCorrAbs} shows that method brings a better measurement of the $\tau$ visible energy.
 
\subsection{Relative correction}

Another strategy to propagate the track energy is to correct the simulation event-by-event using the $E^{cal}/E^{trk}$($=E/p$) distribution for $\Ztautau$ events selected in data:
\begin{equation}
 \left( \frac{E}{p}\right)_{corr} = \left( \frac{E}{p}\right)_{mc} \,\times \, \frac{\langle E/p \rangle_{data}}{\langle E/p \rangle_{mc}}
\end{equation}
Here, $\langle E/p \rangle$ is the average value over the events. Such a method is relative due to the fact that the simulation is adjusted to the data but the true energy is not known either in the data or in the simulation. However, the $R_{\pi}$ measurement is not needed for this approach.

\section{CONCLUSIONS}

In high energy hadronic collisions, $\tau$ leptons require more sophisticated tools than electrons and muons because of their various hadronic decay modes and the escaping neutrino. Such $\tau$ objects are of particular interest to probe the Standard Model as well as physics beyond the SM including many areas such as the electroweak symmetry breaking origin or supersymmetric scenarios. In spite of the various experimental challenges, the \DO experiment has developed an effective algorithm to identify hadronically decaying $\tau$ leptons and has performed several physics measurements and searches employing such technics. In addition, some recent progress has been achieved on the $\tau$ identification improving the potential sensitivity of the experiment with $\tau$ leptons.

\end{document}